\documentclass[journal]{IEEEtran}
\usepackage{amssymb,amsmath}
\usepackage{flushend}
\usepackage{tikz}
\usepackage{pgfplots}
\usepgflibrary{shapes}
\usetikzlibrary{arrows,shapes,chains,matrix,positioning,scopes}

\newtheorem{proposition}{Proposition}

\newlength\tikzheight
\newlength\tikzwidth

\begin{document}

\title{Reconfigurable Antennas, Preemptive Switching and Virtual Channel Management}

\author{\IEEEauthorblockN{Santhosh Kumar,
Jean-Francois Chamberland,
Gregory H. Huff}\\
\thanks{
This publication was made possible by grant NPRP-5-653-2-268 from the Qatar National Research Fund (a member of Qatar Foundation).
The statements made herein are solely the responsibility of the authors.

The authors are with the Department of Electrical and Computer Engineering, Texas A\&M University, College Station, TX 77843, USA (emails: santhosh.kumar@tamu.edu; chmbrlnd@tamu.edu; ghuff@tamu.edu).}
}

\maketitle

\begin{abstract}
This article considers the performance of wireless communication systems that utilize reconfigurable or pattern-dynamic antennas.
The focus is on finite-state channels with memory and performance is assessed in terms of real-time behavior.
In a wireless setting, when a slow fading channel enters a deep fade, the corresponding communication system faces the threat of successive decoding failures at the destination.
Under such circumstances, rapidly getting out of deep fades becomes a priority.
Recent advances in fast reconfigurable antennas provide new means to alter the statistical profile of fading channels and thereby reduce the probability of prolonged fades.
Fast reconfigurable antennas are therefore poised to improve overall performance, especially for delay-sensitive traffic in slow-fading environments.
This potential for enhanced performance motivates this study of the temporal behavior of point-to-point communication systems with reconfigurable antennas.
Specifically, agile wireless communication schemes over erasure channels are analyzed; situations where using reconfigurable antennas yield substantial performance gains in terms of throughput and average delay are identified.
Scenarios where only partial state information is available at the receiver are also examined, naturally leading to partially observable decision processes.
\end{abstract}


\section{Introduction}

High-performance reconfigurable antenna technologies are collectively emerging as a viable option for smartphones, mobile hotspots, and portable computers \cite{huff2004directional,roach2007comparative,yang2009frequency,rajagopalan2009rf,huff2010frequency}.
Reconfigurable antennas can be employed in a number of ways to increase the connectivity profile and robustness of wireless communication channels, and they create new and promising opportunities for the engineering of superior communication schemes.
Such antennas are designed to intentionally and reversibly alter the character of their performance-governing electromagnetic fields.
As a result, they are able to modify the directional and polarization properties of their radiation patterns and thereby change the spatiotemporal characteristics of the communication channels they induce.

Beyond the basic properties of reconfigurable antennas, it is interesting to note that manipulations of radiation patterns can be automated through rapid feedback and triggered by events such as deep fades and successive decoding failures \cite{panagamuwa2006frequency,piazza2008design,mahanfar2008smart,daly2010beamsteering}.
This enables the implementation of closed-loop systems with the cognitive ability to seamlessly adapt to evolving electromagnetic environments and interference conditions.
The capacity to provide these features in (near) real time is primarily determined by the speed and complexity of the reconfiguration mechanisms used to facilitate electromagnetic agility within the reconfigurable antenna structures.
This highlights a natural tradeoff between the potential benefits of a configuration change and the downtime associated with each transformation event.

Although reconfigurable antennas have been and continue to be the subject of concerted research efforts in the antennas and propagation community, a detailed analysis of their repercussions on the foundations of wireless communications is still lacking.
Key to the widespread adoption of such technologies, aside from miniaturization, is provable gains in terms of capacity, delay-throughput profile and network connectivity~\cite{gouaiming}.
This article seeks to address the need for better understanding the impact of adaptive antenna systems.
From a conceptual point of view, fast reconfigurable antennas can be employed to establish ancillary virtual links between two devices.
In a slow fading channel~\cite[p.~31]{TseFWC1111206}, as the quality of the current channel degrades, it may become advantageous to transition to an alternate antenna state and, consequently, to another channel realization.
We envision slow fading scenarios in delay-sensitive applications such as mobile real-time video or gaming in an urban environment. 

This added flexibility at the physical layer is likely to boost the perceived performance of delay-sensitive applications over channels with memory. 
Indeed, several studies document the fact that channel variations are particularly detrimental to delay-constrained communications \cite{Kim2000tnet,Berry2002tit,wu2003effective,Laourine2010twcom}.
Channel memory further exacerbates this situation, as it increases the propensity for prolonged deep fades \cite{Liu2007tit,parag2013code}.
Thus, having the capability to jump to a different virtual channel seems an attractive option in these circumstances.

To carry our analysis of reconfigurable systems, we leverage several results that have appeared in the literature in the past.
First, we adopt a class for channels with memory similar to the finite-state channel model proposed by Gilbert and Elliott~\cite{gilbert1960capacity,elliott1963estimates}.
The delay-sensitive aspect of the problem is captured through a queueing formulation whose solution is obtained, partly, by applying techniques originally developed by Neuts~\cite{Neuts0824782836,Neuts0486683427,hajek1982birth}.
The conceptual bridge between the physical layer and the queueing system is provided by error correcting codes and the availability of feedback.
The framework presented below parallels some of our previous work~\cite{parag2013code,kumar2013firstpassage}.
The incorporation of reconfigurable antennas into the problem setting and the insights provided by our analysis are novel.
This sheds new lights on the potential benefits of adaptive antenna systems and their application to delay-sensitive communication over wireless channels.

Special attention is given to the realistic scenario where the channel state is not known perfectly at the receiver.
Rather, it must be inferred from available observations.
In this latter case, the optimization problem becomes a partially observable decision process.
This forces the controller to make decisions based on an estimated distribution over the channel state.
This problem is cast into a dynamic programming framework with limited state knowledge.
Known system dynamics and partial observations are combined at every step to update the estimated distribution and, subsequently, take action.

The remainder of this article is organized as follows.
The system components, along with a mathematical abstraction for reconfigurable antennas, are described in Section~\ref{section:SystemModel}.
Two modes of operation are considered, a classical system with a static antenna structure and an adaptive implementation with the ability to reconfigure the RF front-end depending on the channel state.
The evolution of these two competing schemes is characterized in Section~\ref{section:QueueingBehavior}, where we analyze their queueing behaviors.
This gives rise to various performance criteria, including throughput, mean waiting time and the probability of the queue exceeding a certain threshold.
Pertinent numerical examples are presented in Section~\ref{section:NumericalResults}.
Finally, concluding remarks and avenues of future research are discussed in the last section.

\section{System Model}
\label{section:SystemModel}

In this study, we consider one side of the communication process.
That is, information flows from a transmitter to a destination.
An eventual assumption on the availability of feedback will necessarily imply the presence of a reverse link.
Nonetheless, for the sake of simplicity, we focus on a single direction with the understanding that a similar analysis can be applied to the reverse link.
We keep the descriptions of the system model to a minimum and refer the reader to \cite{parag2013code,kumar2013firstpassage} for an elaborate exposition.

As mentioned above, overall system performance is assessed using a queueing formulation.
The state of the queue at the source is governed by arrivals and departures.
The evolution of the queue is modeled as a discrete-time stochastic process that is synchronized with codeword transmissions.
Specifically, during each codeword cycle, we assume that a data packet arrives with probability $\gamma$, independently of other time instants.
The number of information bits per data packet, denoted by $L$, is also random and possesses a geometric distribution with parameter $\rho$.
We emphasize that this arrival process is carefully selected to facilitate analysis.
Since this article is primarily concerned with identifying the relative benefits of reconfigurable antennas over traditional implementations, adhering to a specific arrival profile is unlikely to affect the nature of our subsequent results.

\subsection{Traditional Implementation}
\label{section:TraditionalImplementation}

\begin{figure}[tb!]
\begin{center}
  \begin{tikzpicture}
[node distance = 12mm,text height=1.5ex,text depth=.25ex, 
draw=black!80,very thick,
point/.style={coordinate},>=stealth',bend angle=20,
state/.style={circle, inner sep = 0 pt,minimum size = 4mm,thick,draw=black!80,},
Gstate/.style={circle, inner sep = 0pt,minimum size = 8mm,very thick,draw=black!80},
Bstate/.style={circle, inner sep = 0pt,minimum size = 8mm,very thick,draw=black!80},
pre/.style={<-,>=stealth',thick},
post/.style={->,>=stealth',thick}]					

\node[Gstate] (g) at (0,0) {$2$};
  edge[post, bend right=90] (g);
\node[Bstate] (b) [below=of g] {$1$}
  edge[pre, bend right=60] node[auto]{$b_{21}$} (g)
  edge[post, bend left=60] node[auto,swap]{$b_{12}$}(g);
  edge[post, bend right=90] node[auto]{} (b);

\node[point] (gt1)[right=of g, yshift=12,label=left:$1$]{};
\node[point] (gr1)[right=of gt1,,label=right:$1$ ]{}
  edge[pre] (gt1);
\node[point] (gt0)[right=of g, yshift=-12,label=left:$0$]{};
\node[point] (gr0)[right=of gt0,label=right:$0$]{}
  edge[pre](gt0);
\node[point] (gte)[right=of g,label=right:$\varepsilon_2$]{};
\node[point] (gre)[right=of gte,label=right:$e$]{}
  edge[pre] node[auto]{} (gt1)
  edge[pre] (gt0);

\node[point] (bt1)[right=of b, yshift=12,label=left:$1$]{};
\node[point] (br1)[right=of bt1,label=right:$1$]{}
  edge[pre](bt1);
\node[point] (bt0)[right=of b, yshift=-12,label=left:$0$]{};
\node[point] (br0)[right=of bt0,label=right:$0$]{}
  edge[pre](bt0);
\node[point] (bte)[right=of b,label=right:$\varepsilon_1$]{};
\node[point] (bre)[right=of bte,label=right:$e$]{}
  edge[pre] node[auto]{} (bt1)
  edge[pre] (bt0);

\end{tikzpicture}
\caption{A finite-state channel with memory is used to model the operation of a wireless link at the bit level.
For illustrative purposes, the channel is depicted with only two states, a form known as the Gilbert-Elliott channel.}
\label{figure:GilbertElliott}
\end{center}
\end{figure}
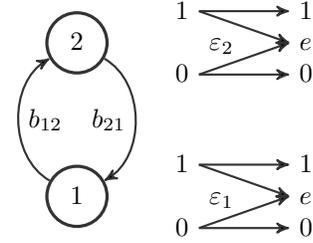

We begin our discussion of the queueing system with a mathematical description of the channel model associated with a fixed antenna configuration.
This communication channel can operate in one of several modes, designated by $\mathcal{C} = \{ 1, \ldots, k \}$.
The evolution of the channel forms a time-homogeneous Markov process and its probability transition matrix is denoted by
\begin{equation} \label{equation:StateTransition}
\mathbf{B} = \begin{bmatrix}
b_{11} & b_{12} & \cdots & b_{1k} \\
b_{21} & b_{22} & \cdots & b_{2k} \\
\vdots & \vdots & \ddots & \vdots \\
b_{k1} & b_{k2} & \cdots & b_{kk} \end{bmatrix} ,
\end{equation}
where $b_{ij}$ symbolizes the probability of jumping to state~$j$ given that the chain is currently in state~$i$.
That is, $\mathbf{B}$ is a right stochastic matrix.
We also assume that the Markov chain governing this finite-state channel is aperiodic and irreducible. 
When in state~$i$, the probability that a bit sent by the source is received faithfully at the destination is equal to $1 - \varepsilon_i$; this bit is consequently erased with probability $\varepsilon_i$.
The elements of the state space $\mathcal{C}$ are indexed in such a way that $i < j$ implies $\varepsilon_i \geq \varepsilon_j$.
In its simplest non-trivial instantiation, this finite-state channel with memory can take on two possible states.
This specific model is known as the Gilbert-Elliott channel~\cite{gilbert1960capacity,elliott1963estimates}, and is illustrated in Fig.~\ref{figure:GilbertElliott}.
Finite-state channels have been employed to model wireless connections in the past, and techniques aimed at selecting representative model parameters are available~\cite{Wang1995tvt,Sadeghi2008spm}.

The adverse effects of channel uncertainty are countered by the communication system using error-control coding.
Data segments are first encoded at the source and then transmitted in the form of codewords to the destination.
As discussed above, individual symbols may or may not be received at the destination depending on the realization of the finite-state channel.
Decoding is executed on a per codeword basis.
We assume that code performance is governed solely by the number of erasures that occur during a transmission interval.
Obtaining a distribution for the number of such erasures, conditioned on the initial state of the channel, is therefore highly desirable.
There exist various strategies to compute such distributions.
Finding the distribution of the channel states and then computing the conditional distribution of the number of erasures is a possible approach~\cite{Wilhelmsson-com99}.
Alternatively, one can employ generating functions and product of matrices with polynomial entries to derive these quantities \cite{parag2013code,kumar2013firstpassage}.
At this point, it suffices to point out that
\begin{equation} \label{equation:NumberOfErasures}
\Pr (E = e, C_{N+1} = j | C_1 = i)
\end{equation}
can be computed efficiently.
Above, $E$ denotes the number of erasures within a block, $C_n$ represents the state of the channel at time $n$, and $N$ denotes the length of a codeword.

We consider a standard and powerful approach to model forward error correction~\cite{Gallager0471290483}.
Every codebook is created using a random binary parity-check matrix $\mathbf{H}$ of size $(N-K) \times N$.
The admissible codewords are the elements of the nullspace of $\mathbf{H}$.
Decoding at the receiver is executed using a maximum likelihood decision rule.
The ensuing probability of decoding failure, conditioned on $e$ erasures, is then given by
\begin{equation} \label{equation:DecodingFailure}
P_{\mathrm{f}}(N-K,e)
= 1 - \prod_{i=0}^{e-1} \left( 1- 2^{i-(N-K)} \right)
\end{equation}
where $N$ is the code length and $K$ designates the number of information bits per codeword~\cite{Richardson0521852293}.
Accounting for channel states, the conditional probability of decoding failure at the destination, which we represent by $P_{\mathrm{df}}(j ; i)$, is equal to
\begin{equation*}
P_{\mathrm{df}}(j ; i)=\sum_{e=0}^N P_{\mathrm{f}}(N-K,e)
\Pr \left( E=e, C_{N+1} = j | C_1 = i \right) .
\end{equation*}
Similarly, the conditional probability of decoding success, labeled $P_{\mathrm{ds}}(j ; i)$, can be written as
\begin{equation*}
\begin{split}
&P_{\mathrm{ds}}(j ; i)\\
&=\sum_{e=0}^N (1 - P_{\mathrm{f}}(N-K,e)) \Pr \left( E=e, C_{N+1} = j | C_1 = i \right) .
\end{split}
\end{equation*}
Thus, combining \eqref{equation:NumberOfErasures} and \eqref{equation:DecodingFailure}, we obtain the probability of transition to state $j$ with or without decoding success, conditioned on the channel being in state $i$.
Collectively, these probabilities underlie the evolution of the queueing system; this is  described next.

To conform with our block encoding scheme, a data packet of length $L$ must be divided into $M = \lceil L / K \rceil$ segments, each of size $K$.
The ending segment of a packet is zero-padded, if needed.
Note that $M$ is too a geometric random variable, albeit with parameter
\begin{equation} \label{equation:RhoR}
\rho_r = \sum_{\ell = 1}^{K} (1 - \rho)^{\ell - 1} \rho
= 1 - (1 - \rho)^{K} .
\end{equation}
These segments are successively encoded into codewords of length $N$ and sent over the finite-state channel.
Upon successful decoding, the destination acknowledges reception of the information and the corresponding segment is discarded from the source buffer.
On the other hand, when transmission fails, the source is notified.
The leading data segment is then immediately re-encoded and transmitted once again over the wireless channel.
This process continues until successful reception of the codeword at the destination.

The number of packets awaiting transmission is selected as the state of the queue.
This perspective reflects our inclination towards delay-sensitive communications.
An alternate formulation would take the number of segments awaiting transmission as the state of the queue.
This latter option would be more appropriate to evaluate the size of the memory necessary to store information at the source, at the expense of a less accurate delay characterization.
It is worth reemphasizing that, in our framework, a packet departs from the transmit buffer whenever a codeword is decoded successfully at the destination and the corresponding segment is the last parcel of information of the lead packet.

Throughout, we use $Q_s$ to identify the state of the queue at discrete-time $s$.
Although the stochastic process $\{ Q_s \}$ does not possess the Markov property, the channel state and the queue length at the onset of a codeword cycle, jointly designated by $Y_s = ( C_{sN + 1}, Q_s)$, form a Markov chain \cite{parag2013code}, \cite[Theorem~1]{kumar2013firstpassage}.
For the sake of completeness, we give a brief argument for this in the following.
The sampled channel process $\{C_{sN+1}\}_{s=0}^\infty$ is clearly Markov since $\{C_n\}$ is a Markov process.
Consider the process $Q_s$.
According to our problem formulation, $Q_{s+1}$ is either $Q_s-1$, $Q_s$ or $Q_{s}+1$ depending on whether the decoding is a success, the segment is the last parcel of the data packet, and there is an arrival.
Conditional on the channel state $C_{sN+1}$, the success of a decoding does not depend on previous channel states and queue levels, but on the generated codebook and the realizations of the channel during the transmission of codeword $s$.
Since the length of a data packet is a geometric random variable, due to its memoryless property, whether the segment is the last parcel of the data packet does not depend on the previous queue transitions.
Moreover, the arrival process is Bernoulli, which is again memoryless.
These observations together imply that the cascaded process $\{(C_{sN+1},Q_s)\}$ is Markovian.

\begin{figure}[tb!]
\begin{center}
  \begin{tikzpicture}
[node distance = 12mm,text height=1.5ex,text depth=.25ex, 
draw=black,very thick,
point/.style={coordinate},>=stealth',bend angle=20,
state/.style={circle, inner sep = 0pt,minimum size = 8mm,very thick,draw=black},
dstate/.style={circle, inner sep = 0pt,minimum size = 8mm,very thick,dashed,draw=black},
pre/.style={<-, >=stealth', thick},
post/.style={->, >=stealth', thick}]					

\node[dstate] (s20) at (0,0) {};
\node[dstate] (s10) [below=of s20] {};
\node[state] (s21) [right=of s20] {\small $2, q$}
  edge[post] node[auto,swap]{\scriptsize $\mu_{22}$} (s20)
  edge[post] node[auto,pos=0.9,swap]{\scriptsize $\mu_{21}$} (s10);
\node[state] (s11) [below=of s21] {\small $1, q$}
  edge[post] node[auto]{\scriptsize $\mu_{11}$} (s10)
  edge[post] node[auto,pos=0.9]{\scriptsize $\mu_{12}$} (s20)
  edge[pre, bend right=11] node[auto,swap]{\scriptsize $\kappa_{21}$} (s21)
  edge[post, bend left=11] node[auto]{\scriptsize $\kappa_{12}$} (s21);
\node[dstate] (s22) [right=of s21] {}
  edge[pre] node[auto,swap]{\scriptsize $\lambda_{22}$} (s21)
  edge[pre] node[auto,pos=0.1]{\scriptsize $\lambda_{12}$} (s11);
\node[dstate] (s12) [below=of s22] {}
  edge[pre] node[auto]{\scriptsize $\lambda_{11}$} (s11)
  edge[pre] node[auto,pos=0.1,swap]{\scriptsize $\lambda_{21}$} (s21);

\end{tikzpicture}
\caption{Possible transitions with partial labeling for a queueing system built upon a two-state channel model and operating using a fixed antenna configuration.
Self-transitions are intentionally omitted.}
\label{figure:TransitionsSet}
\end{center}
\end{figure}
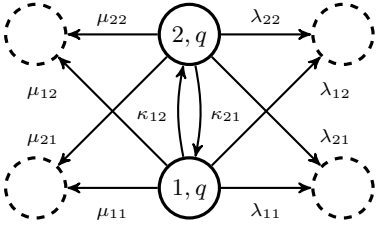

The transition probabilities for this Markov chain can be calculated as follows.
Suppose that the queue is non-empty, i.e., $Y_s = (i, q)$ where $q > 0$.
The admissible values for $Q_{s+1}$ are $\{ q-1, q, q+1 \}$.
Several factors can affect the evolution of the queue over time: the arrival of a new packet, the successful decoding of a codeword and whether or not this latter codeword is the last segment of a data packet.
The only scenario that leads to a decrease in the queue is having no arrival and one packet departure.
Recall that a packet departure occurs when a codeword is successfully decoded and the corresponding segment is the last parcel of information of the lead data packet.
This yields
\begin{equation*}
\begin{split}
\mu_{ij} &= \Pr ( Y_{s+1} = (j, q-1) | Y_s = (i, q) ) \\
&= (1 - \gamma) P_{\mathrm{ds}}(j ; i) \rho_r .
\end{split}
\end{equation*}
For the queue length to remain at a specific level, departures and arrivals must be balanced.
In particular, there can be either no departure and no arrival, or one departure and one arrival,
\begin{equation*}
\begin{split}
\kappa_{ij} &= \Pr ( Y_{s+1} = (j, q) | Y_s = (i, q) ) \\
&= (1 - \gamma) \left( P_{\mathrm{df}}(j ; i)
+ P_{\mathrm{ds}}(j ; i) (1 - \rho_r) \right) \\
&+ \gamma P_{\mathrm{ds}}(j ; i) \rho_r .
\end{split}
\end{equation*}
Finally, the queue length increases whenever a packet arrives and no departure occurs,
\begin{equation*}
\begin{split}
\lambda_{ij} &= \Pr ( Y_{s+1} = (j, q+1) | Y_s = (i, q) ) \\
&= \gamma \left( P_{\mathrm{df}}(j ; i)
+ P_{\mathrm{ds}}(j ; i) (1 - \rho_r) \right) .
\end{split}
\end{equation*}
When the queue is empty, $Q_s = 0$, similar arguments apply, except that there can be no departures,
\begin{align*}
\kappa_{ij}^0 &= \Pr ( Y_{s+1} = (j, 0) | Y_s = (i, 0) ) \\
&= (1 - \gamma) \Pr (C_{(s+1)N+1} = j | C_{sN+1} = i) \\
\lambda_{ij}^0 &= \Pr ( Y_{s+1} = (j, 1) | Y_s = (i, 0) ) \\
&= \gamma \Pr (C_{(s+1)N+1} = j | C_{sN+1} = i) .
\end{align*}
Possible transitions for a non-empty queue at level~$q$ are depicted in Fig~\ref{figure:TransitionsSet}.
Again, for simplicity, the diagram assumes a two-state channel at the physical layer.

\subsection{Adaptive Antenna Implementation}

Although series of carefully designed experiments in anechoic chambers have been reported previously in the literature on reconfigurable antennas~\cite{panagamuwa2006frequency,roach2007comparative,yang2009frequency}, establishing accurate mathematical models for particular system implementations can be a daunting task.
Given that this article only offers a preliminary investigation on the topic, we make simplifying assumptions that are somewhat favorable to RF-agile devices. 
The rationale behind this reasoning is to gain insight without risking to prematurely discard a technology that may eventually lead to significant gains.
We postulate that reconfigurable antennas have a large number of possible configurations, and we assume that the wireless channels induced by these configurations are independent from one another.
This is more likely to apply to situations where devices are embedded in rich scattering environments.
A direct implication of these two hypotheses is that a wireless device equipped with a reconfigurable antenna can always elect to switch to a different virtual channel.
Furthermore, once this transformation is accomplished, the probability that the wireless channel occupies a particular state becomes equal to the stationary probability of this same state.

The second aspect of reconfigurable antennas that warrants attention is the latency of the morphing process.
The amount of time necessary to execute an antenna reconfiguration depends heavily on the physics underlying the process.
Envisioned technologies related to our present investigation include electronic switches, microelectromechanical systems (MEMS) and microfluidic devices.
Collectively, these various techniques embody a range of options in terms of latency, efficiency and power consumption.
They also offer fundamentally different mechanisms that can provide measurable tradeoffs between speed, power handling, linearity and overall complexity.
Moreover, they each feature a compact form factor suitable for mobile devices.
Based on the state-of-the-art for these mechanisms, it is reasonable to assume that reconfiguration latency is no greater than a typical codeword cycle (on the order of 4.615~ms).
In our analysis, we assume that triggering an antenna reconfiguration event results in the loss of one codeword transmission opportunity;
no segment can be decoded at the completion of the corresponding transmission interval and, as such, there cannot be a departure from the queue.
This is the price to pay for the opportunity to access a fresh channel realization.
We note that, although not pursued in this article, it is possible to extend the ensuing analysis to scenarios where there is a loss of more than one codeword transmission opportunity per reconfiguration event.

We consider static control policies for antenna handling that are based solely on channel state.
Furthermore, we look at hierarchical structures: if channel state $i$ is deemed deficient enough to initiate a channel reconfiguration, then channel state $j$ will also trigger a reconfiguration whenever $j \leq i$.
While it may be possible to create idiosyncratic channel probability transition matrices for which this class of policies is highly suboptimal, hierarchical structures are expected to work well for realistic channel models derived from empirical observations.
Implicit to these control policies is channel state knowledge at the source; this construction again favors adaptive systems.
More pragmatic schemes would have to employ state estimates or trigger a reconfiguration based on the number of successive failed decoding attempts.
In Section-\ref{subsection:POMDPPerspective}, we consider policies based on the estimated channel state distribution and compare the performance of the two competing implementations.

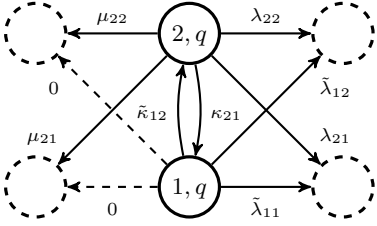
\begin{figure}[tb!]
\begin{center}
  \begin{tikzpicture}
[node distance = 12mm,text height=1.5ex,text depth=.25ex, 
draw=black,very thick,
point/.style={coordinate},>=stealth',bend angle=20,
state/.style={circle, inner sep = 0pt,minimum size = 8mm,very thick,draw=black},
dstate/.style={circle, inner sep = 0pt,minimum size = 8mm,very thick,dashed,draw=black},
pre/.style={<-, >=stealth', thick},
post/.style={->, >=stealth', thick}]					

\node[dstate] (s20) at (0,0) {};
\node[dstate] (s10) [below=of s20] {};
\node[state] (s21) [right=of s20] {\small $2, q$}
  edge[post] node[auto,swap]{\scriptsize $\mu_{22}$} (s20)
  edge[post] node[auto,pos=0.9,swap]{\scriptsize $\mu_{21}$} (s10);
\node[state] (s11) [below=of s21] {\small $1, q$}
  edge[post,dashed] node[auto]{\scriptsize $0$} (s10)
  edge[post,dashed] node[auto,pos=0.9]{\scriptsize $0$} (s20)
  edge[pre, bend right=11] node[auto,swap]{\scriptsize $\kappa_{21}$} (s21)
  edge[post, bend left=11] node[auto]{\scriptsize $\tilde{\kappa}_{12}$} (s21);
\node[dstate] (s22) [right=of s21] {}
  edge[pre] node[auto,swap]{\scriptsize $\lambda_{22}$} (s21)
  edge[pre] node[auto,pos=0.1]{\scriptsize $\tilde{\lambda}_{12}$} (s11);
\node[dstate] (s12) [below=of s22] {}
  edge[pre] node[auto]{\scriptsize $\tilde{\lambda}_{11}$} (s11)
  edge[pre] node[auto,pos=0.1,swap]{\scriptsize $\lambda_{21}$} (s21);

\end{tikzpicture}
\caption{In this diagram, the two-state antenna system evolves unaltered while in state $(c_2, q)$; whereas an antenna reconfiguration is initiated whenever the system enters state $(c_1, q)$.
The reconfiguration process alters the transition probability of the system, as designated by the tildes.}
\label{figure:TransitionsReconf}
\end{center}
\end{figure}

When the channel state is deemed satisfactory, no reconfiguration takes place and the transition probabilities defined in Section~\ref{section:TraditionalImplementation} apply.
On the other hand, when a system reconfiguration is initiated, the transition probabilities become
\begin{align*}
\tilde{\mu}_{ij} &= \Pr ( Y_{s+1} = (j, q-1) | Y_s = (i, q) ) = 0 \\
\tilde{\kappa}_{ij} &= \Pr ( Y_{s+1} = (j, q) | Y_s = (i, q) )
= (1 - \gamma) p_C (j) \\
\tilde{\lambda}_{ij} &= \Pr ( Y_{s+1} = (j, q+1) | Y_s = (i, q) )
= \gamma p_C (j) ,
\end{align*}
where $p_C (\cdot)$ is the marginal probability distribution of the individual wireless channels.
We emphasize that these probabilities are completely determined by the latter stationary distribution and the probability of a packet arrival.
In a similar fashion, we have
\begin{xalignat*}{2}
\tilde{\kappa}_{ij}^0 &= (1 - \gamma) p_C (j) &
\tilde{\lambda}_{ij}^0 &= \gamma p_C (j)
\end{xalignat*}
whenever a reconfiguration event is sparked from an empty queue.
Figure~\ref{figure:TransitionsReconf} presents a level-transition diagram for a two-state channel where an antenna reconfiguration is prompted whenever the channel state lies in the lowest level at the onset of a codeword cycle.
It may be instructive to compare this graph with Fig.~\ref{figure:TransitionsSet}, whose labels embody the operation of a communication system with a static antenna structure.

This completes the description of the queueing systems, with and without reconfigurable antenna structures.
In both cases, the state space for the discrete-time packetized system is $\mathcal{C} \times \mathbb{N}_0$.
Each implementation will be stable provided that the average arrival rate is less than its expected service rate.
When this is the case, the underlying Markov chain is positive recurrent and it admits a stationary distribution~\cite{Bremaud0387985093}.
In the next section, we further discuss stability conditions and we provide means to compute invariant distributions.
This is performed by linking the mathematical formulation of our problem to classical queueing results.

\section{Queueing Behavior}
\label{section:QueueingBehavior}

Recall that a new data packet arrives at the source at time~$s$ with probability~$\gamma$.
Moreover, the number of segments contained in any packet is a geometric random variable with parameter $\rho_r$, as described in \eqref{equation:RhoR}.
The expected arrival rate in segments per block is then given by
\begin{equation*}
\gamma \mathrm{E} \left[ M \right]
= \frac{\gamma}{\rho_r} .
\end{equation*}

The expected service rate depends on the communication scheme employed.
In the traditional implementation with static antennas, the progression of the wireless channel is unaltered at the codeword boundaries.
The throughput in segments per block can be expressed as
\begin{equation}
\label{equation:ServiceRateSet}
\sum_{i \in \mathcal{C}} \sum_{j \in \mathcal{C}}
P_{\mathrm{ds}} (j ; i) p_C (i) ,
\end{equation}
where $p_C (\cdot)$ represents the stationary channel distribution associated with matrix
\begin{equation} \label{equation:TransitionMatrixN}
\mathbf{B}^N = \begin{bmatrix}
b_{11}^{(N)} & b_{12}^{(N)} & \cdots & b_{1k}^{(N)} \\
b_{21}^{(N)} & b_{22}^{(N)} & \cdots & b_{2k}^{(N)} \\
\vdots & \vdots & \ddots & \vdots \\
b_{k1}^{(N)} & b_{k2}^{(N)} & \cdots & b_{kk}^{(N)}
\end{bmatrix} .
\end{equation}
We stress that, since $\mathbf{B}$ is assumed irreducible and aperiodic, its invariant distribution $p_C(\cdot)$ exists and is unique~\cite{Norris0521633966}.
This distribution is also invariant for probability transition matrix $\mathbf{B}^N$, which justifies its use in \eqref{equation:ServiceRateSet}.

Obtaining the probability transition matrix for the adaptive architecture with reconfigurable antennas is a slightly more involved task.
Let $\mathcal{C}^{\dagger} = \{ \ell, \ldots, k \}$ represent the collection of channel states judged suitable for data transmission.
Then, necessarily, the set $\mathcal{C} \setminus \mathcal{C}^{\dagger}$ contains all the channel states for which a reconfiguration command is issued.
In view of this partitioning, we gather that the probability transition matrix for the channel state at the onset of a codeword cycle is
\begin{equation}
\tilde{\mathbf{B}}^{(N)} = \begin{bmatrix}
p_C(1) & p_C(2) & \cdots & p_C(k) \\
\vdots & \vdots & \ddots & \vdots \\
p_C(1) & p_C(2) & \cdots & p_C(k) \\
b_{\ell 1}^{(N)} & b_{\ell 2}^{(N)} & \cdots & b_{\ell k}^{(N)} \\
\vdots & \vdots & \ddots & \vdots \\
b_{k1}^{(N)} & b_{k2}^{(N)} & \cdots & b_{kk}^{(N)}
\end{bmatrix} .
\end{equation}
We point out that matrix entry $b_{ij}^{(N)}$ is implicitly defined in \eqref{equation:TransitionMatrixN}.
Given that successful decoding is only possible when a codeword is sent, we can write the throughput for the adaptive system as
\begin{equation*}
\sum_{i \in \mathcal{C}^{\dagger}} \sum_{j \in \mathcal{C}}
P_{\mathrm{ds}} (j ; i) \tilde{p}_C (i) ,
\end{equation*}
where $\tilde{p}_C (\cdot)$ is the invariant distribution associated with probability transition matrix $\tilde{\mathbf{B}}^{(N)}$.

When the average arrival rate is strictly less than the expected service rate, Foster's criteria guarantees that the corresponding Markov chain is positive recurrent~\cite[p.~167]{Bremaud0387985093}.
It is instructive to point out that channel memory and channel quality can greatly influence the expected service rate of a communication system.
This phenomenon is rooted in the subtle interactions between channel output sequences and the probability of decoding success at the destination.
This is illustrated through numerical examples in the next section.

The channel state and the queue length at the onset of a codeword cycle, $Y_s = (C_{sN+1}, Q_s)$, jointly form a stochastic process with a countably infinite state space.
A natural ordering for its elements is the following,
\begin{equation*}
(1, 0), \ldots, (k, 0),
(1, 1), \ldots, (k, 1),
(1, 2), \ldots
\end{equation*}
Collectively, the subset of states
\begin{equation*}
\left\{ (1, q), \ldots, (k, q) \right\}
\end{equation*}
is known as the $q$th level of the chain.
Using this ordering and the level abstraction, we can introduce a probability transition operator $\mathbf{T}$ for aggregate chain $\{ Y_s \}$,
\begin{equation} \label{equation:OperatorT}
\mathbf{T} = \left( \begin{array}{ccccc}
\mathbf{C}_1 & \mathbf{C}_2 & \mathbf{0} & \mathbf{0} & \cdots \\
\mathbf{A}_0 & \mathbf{A}_1 & \mathbf{A}_2 & \mathbf{0} & \cdots \\
\mathbf{0} & \mathbf{A}_0 & \mathbf{A}_1 & \mathbf{A}_2 & \cdots \\
\mathbf{0} & \mathbf{0} & \mathbf{A}_0 & \mathbf{A}_1 & \cdots \\
\vdots & \vdots & \vdots & \vdots & \ddots
\end{array} \right) .
\end{equation}
In the case of a fixed antenna configuration, the submatrices $\mathbf{A}_0, \mathbf{A}_1, \mathbf{A}_2$ are given by
\begin{xalignat*}{2}
\mathbf{A}_0 &= \begin{bmatrix} \mu_{11} & \cdots & \mu_{1k} \\
\vdots & \ddots & \vdots \\
\mu_{k1} & \cdots & \mu_{kk} \end{bmatrix} &
\mathbf{A}_1 &= \begin{bmatrix} \kappa_{11} & \cdots & \kappa_{1k} \\
\vdots & \ddots & \vdots \\
\kappa_{k1} & \cdots & \kappa_{kk} \end{bmatrix} \\
\mathbf{A}_2 &= \begin{bmatrix} \lambda_{11} & \cdots & \lambda_{1k} \\
\vdots & \ddots & \vdots \\
\lambda_{k1} & \cdots & \lambda_{kk} \end{bmatrix} .
\end{xalignat*}
When the queue is empty, the block transitions are governed by matrices
\begin{xalignat*}{2}
\mathbf{C}_1 &= \begin{bmatrix} \kappa_{11}^0 & \cdots & \kappa_{1k}^0 \\
\vdots & \ddots & \vdots \\
\kappa_{k1}^0 & \cdots & \kappa_{kk}^0 \end{bmatrix} &
\mathbf{C}_2 &= \begin{bmatrix} \lambda_{11}^0 & \cdots & \lambda_{1k}^0 \\
\vdots & \ddots & \vdots \\
\lambda_{k1}^0 & \cdots & \lambda_{kk}^0 \end{bmatrix} .
\end{xalignat*}
These definitions have analog counterparts for systems with reconfigurable antennas; the appropriate entries are simply replaced by their homologues,
\begin{equation*}
\tilde{\mu}_{ij},~\tilde{\kappa}_{ij},~\tilde{\lambda}_{ij},
~\tilde{\kappa}_{ij}^0 \text{ and } \tilde{\lambda}_{ij}^0.
\end{equation*}
In both scenarios, with and without adaptation, the ensuing Markov chains belong to the class of random processes with repetitive structures~\cite{Neuts0486683427}.
One can find the stationary distribution associated with $\mathbf{T}$ by inspecting the substochastic matrix $\mathbf{U}$ whose entry $u_{ij}$ denotes the probability that, starting from state $(1, i)$, the Markov chain $\{ Y_s \}$ first re-enters level one through $(1,j)$ and does so without visiting any state at level zero.
A probabilistic path-counting argument leads to Proposition~\ref{proposition:TabooProbability} \cite[Theorem~4.2]{parag2013code}.

\begin{proposition} \label{proposition:TabooProbability}
Define $\mathbf{U}_1 = \mathbf{A}_1$.
The iterative expression
\begin{equation*}
\mathbf{U}_{m+1} =
\mathbf{A}_1 + \mathbf{A}_2 \left( \mathbf{I} - \mathbf{U}_m \right)^{-1} \mathbf{A}_0
\end{equation*}
is well-defined for all $m \in \mathbb{N}$; its limit exists and
\begin{equation*}
\lim_{m \rightarrow \infty} \mathbf{U}_m
= \mathbf{U} .
\end{equation*}
\end{proposition}

Let $\pi$ represent the invariant distribution of the augmented Markov chain, and denote the subcomponents associated with level~$q$ by $\pi_q$,
\begin{equation*}
\pi_q = \left( \Pr(Y = (1, q)), \ldots, \Pr (Y = (k, q)) \right) .
\end{equation*}

\begin{proposition} \label{proposition:InvariantDistribution}
Define $\mathbf{R} = \mathbf{A}_2 (\mathbf{I} - \mathbf{U})^{-1}$ and recall that the entries of $\pi$ are non-negative and sum up to one.
The invariant distribution induced by $\mathbf{T}$ is entirely determined through the following relations,
\begin{equation} \label{equation:ReducedMatrix}
\begin{bmatrix} \pi_0 & \pi_1 \end{bmatrix}
\begin{bmatrix} \mathbf{C}_1 & \mathbf{C}_2 \\
\mathbf{A}_0 & \mathbf{A}_1 + \mathbf{R} \mathbf{A}_0 \end{bmatrix}
= \begin{bmatrix} \pi_0 & \pi_1 \end{bmatrix}
\end{equation}
and $\pi_q = \pi_1 \mathbf{R}^{q-1}$ for $q \geq 1$.
\end{proposition}

Propositions~\ref{proposition:TabooProbability} and~\ref{proposition:InvariantDistribution} provide an algorithmic blueprint to compute the stationary distribution of the augmented Markov chain:
obtain $\mathbf{U}$ through repeated iterations;
compute $\mathbf{R}$ and form irreducible and aperiodic probability transition matrix
\begin{equation*} 
\begin{bmatrix} \mathbf{C}_1 & \mathbf{C}_2 \\
\mathbf{A}_0 & \mathbf{A}_1 + \mathbf{R} \mathbf{A}_0 \end{bmatrix} ;
\end{equation*}
find its invariant distribution;
append missing values of $\pi$ using $\pi_q = \pi_1 \mathbf{R}^{q-1}$ and normalize.

\section{Performance Evaluation} 
\label{section:PerformanceEvaluation}

Once the stationary distribution is acquired, we can compute several performance criteria of interest.
In this article, we examine the average delay, the probability that the queue length exceeds a certain threshold and the decay rate of the queue occupancy.
First, we note that the expected queue length is given by
\begin{equation} \label{equation:MeanQueueLength}
\sum_{q=0}^{\infty} q \pi_q \cdot \mathbf{1}
= \pi_1 \left( \sum_{q=1}^{\infty}
q \mathbf{R}^{q-1} \right) \cdot \mathbf{1} ,
\end{equation}
where $\mathbf{1}$ is a column vector of all ones.
Using Little's formula, we deduce that the mean waiting time in the queue is simply \eqref{equation:MeanQueueLength} divided by expected arrival rate $\gamma$
\begin{equation*}
\frac{1}{\gamma} \pi_1 \left( \sum_{q=1}^{\infty}
q \mathbf{R}^{q-1} \right) \cdot \mathbf{1} .
\end{equation*}
The decay rate of the queue occupancy can be written as
\begin{equation} \label{equation:taildecay}
\lim_{\tau \rightarrow \infty} \frac{1}{\tau} \log \Pr(Q \geq \tau)
= \log \varrho(\mathbf{R}) ,
\end{equation}
where $\varrho(\mathbf{R})$ is the spectral radius of $\mathbf{R}$;
and its complementary cumulative distribution function is determined by the finite sum
\begin{equation*}
\Pr(Q > \tau)
= 1 - \sum_{q=0}^{\lfloor \tau \rfloor} \pi_q \cdot \mathbf{1} .
\end{equation*}

\subsection{Partially Observable Systems}
\label{subsection:POMDPPerspective}

Up to this point, we have looked at the possible benefits of reconfigurable antenna systems by comparing the performance of specific policies using various optimization objectives, e.g., throughput, average delay, queue threshold violation.
To offer a different perspective, we study the communication process in a decision theoretic framework in which the exact state of the channel is unknown.
At the onset of every codeword, the transmitter has the freedom to select an appropriate code-rate or it can trigger an antenna reconfiguration when the channel quality is deemed deficient.
The exact state of the channel is hidden from the transmitter, yet the device can estimate it from the acknowledgements of previous transmission attempts.
In this latter scenario, the objective is to maximize the amount of information transmitted in the long run.
The formulation of a \emph{partially observable Markov decision process} (POMDP) is immediate.
A distinguishing feature of the proposed framework is that it naturally incorporates state estimation, code rate selection, antenna reconfiguration and throughput maximization.

We note that the transmitter must base its actions on past observations.
To act optimally, it is sufficient, surprisingly, to summarize previous observations into a \emph{belief state} \cite{smallwood1973optimal}.
One such choice for the belief state is the probability distribution of the channel state conditioned on the past observations.
Whenever the transmitter takes an action and collects an observation, this belief state must be updated.
We refer the reader to \cite{kaelbling_planning_1998} for the notation and solution methodologies for POMDPs.
Below, we formally define the quantities involved in our problem formulation. 

A partially-observable Markov decision process consists of a tuple $(\mathcal{C}, \mathcal{B}, \mathcal{A}, \mathcal{O}, \mathcal{T}, \mathcal{R}, \Omega)$, as described below.
\begin{itemize}
\item As before, $\mathcal{C}$ represents the set of admissible channel states, $\{ 1, \ldots, k \}$, and $C_n$ denotes the realization of the channel at time instant $n$.
\item $\mathcal{B} \subset [0,1]^{k}$ is the collection of prior distributions (or \emph{belief states}) on the channel state at the transmitter.
The belief at time instant $n$ is written as $B_n$.
\item $\mathcal{A}$ is the space of actions, which includes different code rates and the reconfiguration event.
Code rates of the form $a/N$, $1 \leq a \leq N$, $a \in \mathbb{N}$ are valid options.
We index the code-rate $a/N$ with $a$, and a reconfiguration event is labeled by $0$.
The action at time instant $n$ is denoted by $A_n$.
\item $\mathcal{O}$ denotes the set of observations; these assume the form of an ACK or NACK obtained form the receiver.
The observation at time instant $n$ is denoted by $O_n$.
  \item $\Omega : \mathcal{B} \times \mathcal{A} \times \mathcal{O} \rightarrow \mathbb{R}^{+}$ is the observation function which gives, for each action and a belief state, the probability of making an observation.
\item $\mathcal{T}: \mathcal{B} \times \mathcal{A} \times \mathcal{O} \rightarrow \mathcal{B}$ denotes the transition function which produces, for each action and an observation, the updated belief state as a function of the previous belief state.
\item $\mathcal{R}: \mathcal{B} \times \mathcal{A} \rightarrow \mathbb{R}^{+}$ gives the reward received for the corresponding action.
For our problem, this is the number of transmitted information bits, if any, normalized by the length of a codeword and weighted by the probability of the successful reception.
\end{itemize}

Having established a proper framework, we proceed with the computation of the quantities $\Omega$, $\mathcal{T}$, $\mathcal{R}$  for our problem.
By definition, the observation function is given by
\begin{equation*}
\Omega(\psi, a, o) = \Pr (O_{n+1}=o | B_n=\psi, A_n=a) ,
\end{equation*}
and, hence,
\begin{equation*}
\Omega(\psi, a, o)
= \begin{cases} \psi \mathbf{P}_{\mathrm{ds}} \mathbf{1}, & o = \mathrm{ACK} \\
\psi \mathbf{P}_{\mathrm{df}} \mathbf{1}, & o = \mathrm{NACK} \end{cases}
\end{equation*}
for $a \neq 0$.
The matrices $\mathbf{P}_{\mathrm{ds}}$ and $\mathbf{P}_{\mathrm{df}}$ are given entrywise by
\begin{align*}
[\mathbf{P}_{\mathrm{ds}}]_{ij} &= P_{\mathrm{ds}}(j;i)
& [\mathbf{P}_{\mathrm{df}}]_{ij} &= 1-P_{\mathrm{ds}}(j;i) .
\end{align*}
When a reconfiguration request is issued, we invariably get $\Omega(\psi, 0, \mathrm{ACK}) = 0$ and $\Omega(\psi, 0, \mathrm{NACK}) = 1$.
The transfer function $\mathcal{T}$ updates the current belief state based on the action taken and the observation made.
That is, 
\begin{align*}
  [\mathcal{T}(\psi,a,o)]_{i}=\Pr \left( C_{n+1}=i | B_n=\psi, O_n=o, A_n=a \right) .
\end{align*}
This can be reduced to
\begin{align*}
[\mathcal{T}(\psi,a,o)]_{i}=\frac{\Pr \left( C_{n+1}=i, O_n=o | A_n=a, B_n=\psi \right)}{\Pr \left( O_n=o | A_n=a, B_n=\psi \right)} .
\end{align*}
Writing the above equation compactly, for $a \neq 0$, we get
\begin{align*}
\mathcal{T}(\psi,a,\mathrm{ACK})
&= \frac{\psi \mathbf{P}_{\mathrm{ds}}}{\psi \mathbf{P}_{\mathrm{ds}} \mathbf{1}}
&\mathcal{T}(\psi,a,\mathrm{NACK})
&= \frac{\psi \mathbf{P}_{\mathrm{df}}}{\psi \mathbf{P}_{\mathrm{df}} \mathbf{1}} .
\end{align*}
For the reconfiguration event, which we assume leads to a new virtual channel, we obtain
\begin{equation*}
\mathcal{T}(\psi,0,\mathrm{ACK})
= \mathcal{T}(\psi,0,\mathrm{NACK}) = p_{C} .
\end{equation*}
The reward function is the expected number of information bits received at the destination under state $\psi$ upon action $a$, normalized by the length of a codeword,
\begin{equation*}
\mathcal{R} (\psi, a) = \tfrac{a}{N} \psi \mathbf{P}_{\mathrm{ds}} \mathbf{1} .
\end{equation*}

We define an $n$-step policy $\Delta_n=\{\delta_1,\dots,\delta_n\}$ as a collection of $n$ functions that map each belief state in $\mathcal{B}$ to an action from $\mathcal{A}$. 
The expected discounted reward function $V_{\Delta_n,n}:\mathcal{B}\rightarrow \mathbb{R}^{+}$ is given by
\begin{align*}
  V_{\Delta_n,n}(\psi) &= \mathrm{E} \left[ \sum_{t=1}^{n} \beta^{t-1} \mathcal{R} (B_t,\delta_t(B_t)) \right] ,
\end{align*}
where $B_1=\psi$.
Intuitively, $V_{\Delta_n,n}$ denotes the expected rewards received by executing the policy $\Delta_n$ for $n$-steps.
However, the rewards procured in the future are discounted by a factor of $\beta < 1$ each time.
In a delay-sensitive system, the discount factor can be thought of as penalizing the information bits that will be transmitted in the future.
Still, this is a standard approach that ensures convergence and the existence of a solution.
The $n$-step finite horizon value function is defined as the maximum of $V_{\Delta_n,n}$ over all the policies,
\begin{align*}
  V_{n} = \max_{\Delta_n} V_{\Delta_n,n} .
\end{align*}
By the \emph{principle of optimality}, the finite horizon value functions satisfy~\cite{bertsekas2005dynamic}
\begin{equation*}
\begin{split}
V_{n}(\psi)
= \max_{a \in \mathcal{A}} \Big[ & \mathcal{R}(\psi,a) \\
&+ \beta \sum \limits_{o\in\mathcal{O}} \Omega(o,\psi,a)
V_{n-1}(\mathcal{T}(\psi,o,a)) \Big] .
\end{split}
\end{equation*}
Similarly, the infinite horizon value function is defined as
\begin{equation*}
V(\psi) = \max_{\Delta} \mathrm{E} \left[ \sum_{t=1}^{\infty} \beta^{t-1} \mathcal{R} (B_t,\delta_t(B_t)) \right] ,
\end{equation*}
where $B_1=\psi$ and $\Delta$ is a sequence of functions $\{\delta_t\}_{t=1}^{\infty}$ which map each belief state to an action.
The optimal policy $\Delta^{*}$ that maximizes the above equation is a stationary policy~\cite{sondik1978optimal}.
That is $\Delta^{*}=\{ \delta_t=\delta^{*} \}_{t=1}^{\infty}$ for some function $\delta^{*}: \mathcal{B} \rightarrow \mathcal{A}$.   
The infinite horizon value function can be approximated by a finite horizon value function according to~\cite[Theorem~3]{sondik1978optimal}
\begin{equation*}
V(\psi) = \lim_{n \rightarrow \infty} V_n(\psi) .
\end{equation*}

In the next section, we use these performance criteria to show that time-dependencies in the underlying physical channel can adversely affect the behavior of a queueing system.
Moreover, having the ability to reconfigure an antenna structure at appropriate moments can help mitigate the undesirable effects of channel memory.

\section{Numerical Results}
\label{section:NumericalResults}

Although the methodology introduced above applies to a broad class of finite-state erasure channels with memory, our numerical study focuses mostly on the two-state Gilbert-Elliott model depicted in Fig.~\ref{figure:GilbertElliott}, and finite-state channels derived from quantizing Rayleigh fading channels \cite{Wang1995tvt,Zhang1999tcom}.
These models capture many of the features associated with wireless environments such as uncertainty, fading and channel memory.
Yet, they remain simple enough for a straightforward exposition of our findings.
Overall, these models provide valuable insights about the operation of wireless communication systems without being overly intricate, which gives ground for their adoption.
This is especially relevant for a first characterization of the potential benefits associated with reconfigurable antenna structures.

\subsection{Gilbert-Elliott Model}

This channel model captures the bursty nature of a wireless environment and has received a fair amount of attention in the literature.
The choice of a hierarchical control policy for the Gilbert-Elliott channel with side information is straightforward.
The only non-trivial candidate is the adaptive policy where the source triggers an antenna reconfiguration whenever the state of the channel is $1$ at the onset of a codeword transmission.
The performance of this adaptive scheme is compared with the operation of a static system where the antenna structure is fixed and codewords are sent at every opportunity.

For the Gilbert-Elliott channel model, the stochastic matrix $\mathbf{B}$ reduces to
\begin{equation*}
\mathbf{B} = \left[ \begin{array}{cc} b_{11} & b_{12} \\
b_{21} & b_{22} \end{array} \right] ,
\end{equation*}
and it has only two degrees of freedom.
One possible way to portray these degrees of freedom is to talk about the stationary probabilities of the states,
\begin{xalignat*}{2}
\Pr (C = 1) &= \frac{b_{21}}{b_{12} + b_{21}} &
\Pr (C = 2) &= \frac{b_{12}}{b_{12} + b_{21}} ,
\end{xalignat*}
and channel memory;
this is the approach we use throughout.
The memory of a Gilbert-Elliott channel can be expressed at the symbol level using $1 - b_{12} - b_{21}$.
An equivalent way to characterize memory is to consider changes at the codeword level,
\begin{equation} \label{equation:ChannelMemoryParameter}
1 - b_{12}^{(N)} - b_{21}^{(N)}
= \left( 1 - b_{12} - b_{21} \right)^N
\in [0, 1) .
\end{equation}
It is typically more insightful to plot results using the latter scaling and, as such, this is the unit we employ in our figures.
Additional system parameters are selected to approximate the operation of a typical communication link.
The block length is set to $N = 114$.
New packets arrive at the source with probability $\gamma = 0.20$, and their expected length is $\rho^{-1} = 195$ bits.
Based on a 4.615~ms codeword cycle, this yields a lightly loaded connection at roughly 8.45~kbps;
these are realistic numbers for digital telephony.

We assess the performance of our competing systems when operating over erasure channels with an erasure probability equal to 20 percent.
We first explore the impact of channel memory on overall performance.
We study a channel model with $b_{12} = 4 b_{21}$, $\varepsilon_1 = 0.5$ and $\varepsilon_2 = 0.125$.
Channel correlation over time is varied progressively from the memoryless case to a very slow fading profile.
Note that the value of $K$ is throughput optimized for every parameter set and system implementation, leading to a fair comparison between schemes.

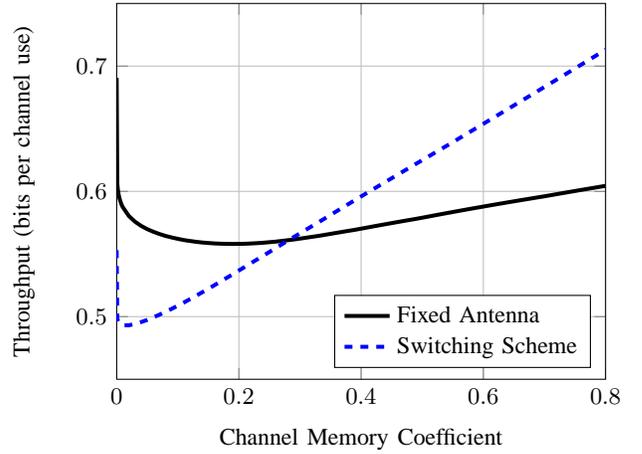
\begin{figure}[tb!]
\begin{center}
\setlength\tikzheight{5cm}
\setlength\tikzwidth{6.5cm}
%
%
\begin{tikzpicture}

\begin{axis}[%
font=\small,
scale only axis,
width=\tikzwidth,
height=\tikzheight,
xmin=0, xmax=0.8,
ymin=0.45, ymax=0.75,
xlabel={Channel Memory Coefficient},
ylabel={Throughput (bits per channel use)},
xmajorgrids,
ymajorgrids,
legend entries={Fixed Antenna, Switching Scheme},
legend style={at={(0.97,0.03)},anchor=south east,nodes=right}]
\addplot [
color=black,
solid,
line width=1.5pt
]
coordinates{ (0,0.690556) (0.001,0.606174) (0.002,0.601228) (0.003,0.598071) (0.004,0.595705) (0.005,0.593793) (0.006,0.59218) (0.007,0.590779) (0.008,0.589538) (0.009,0.588422) (0.01,0.587406) (0.01,0.587406) (0.02,0.580301) (0.03,0.575799) (0.04,0.572451) (0.05,0.569843) (0.06,0.567753) (0.07,0.565966) (0.08,0.564415) (0.09,0.56319) (0.1,0.5621) (0.11,0.56112) (0.12,0.560389) (0.13,0.55973) (0.14,0.559166) (0.15,0.558776) (0.16,0.558427) (0.17,0.558213) (0.18,0.558086) (0.19,0.558009) (0.2,0.558083) (0.21,0.55817) (0.22,0.558396) (0.23,0.558661) (0.24,0.558994) (0.25,0.559427) (0.26,0.55986) (0.27,0.560418) (0.28,0.561005) (0.29,0.561587) (0.3,0.562279) (0.31,0.563003) (0.32,0.563719) (0.33,0.564427) (0.34,0.565235) (0.35,0.566072) (0.36,0.566899) (0.37,0.567717) (0.38,0.568526) (0.39,0.569325) (0.4,0.570213) (0.41,0.571125) (0.42,0.572027) (0.43,0.572918) (0.44,0.5738) (0.45,0.574673) (0.46,0.575537) (0.47,0.576391) (0.48,0.577237) (0.49,0.578076) (0.5,0.57896) (0.51,0.57989) (0.52,0.58081) (0.53,0.581722) (0.54,0.582625) (0.55,0.583519) (0.56,0.584405) (0.57,0.585283) (0.58,0.586153) (0.59,0.587016) (0.6,0.587871) (0.61,0.588719) (0.62,0.58956) (0.63,0.590394) (0.64,0.591222) (0.65,0.592043) (0.66,0.592858) (0.67,0.593666) (0.68,0.594468) (0.69,0.595265) (0.7,0.596055) (0.71,0.59691) (0.72,0.597766) (0.73,0.598615) (0.74,0.599458) (0.75,0.600295) (0.76,0.601126) (0.77,0.601952) (0.78,0.602771) (0.79,0.603585) (0.8,0.604394)
};

\addplot [
color=blue,
dashed,
line width=1.5pt
]
coordinates{ (0,0.553042) (0.001,0.498921) (0.002,0.496881) (0.003,0.495747) (0.004,0.494995) (0.005,0.494455) (0.006,0.49405) (0.007,0.493739) (0.008,0.493498) (0.009,0.493309) (0.01,0.493162) (0.01,0.493162) (0.02,0.493099) (0.03,0.494086) (0.04,0.49564) (0.05,0.497375) (0.06,0.499435) (0.07,0.501534) (0.08,0.50387) (0.09,0.506213) (0.1,0.508748) (0.11,0.511267) (0.12,0.513967) (0.13,0.516632) (0.14,0.519446) (0.15,0.522255) (0.16,0.525082) (0.17,0.528023) (0.18,0.530929) (0.19,0.533803) (0.2,0.536807) (0.21,0.539801) (0.22,0.542766) (0.23,0.545704) (0.24,0.548617) (0.25,0.551628) (0.26,0.55465) (0.27,0.557649) (0.28,0.560626) (0.29,0.563584) (0.3,0.566525) (0.31,0.569448) (0.32,0.572357) (0.33,0.575252) (0.34,0.578205) (0.35,0.581192) (0.36,0.584166) (0.37,0.587128) (0.38,0.590079) (0.39,0.593019) (0.4,0.59595) (0.41,0.598872) (0.42,0.601786) (0.43,0.604693) (0.44,0.607594) (0.45,0.610489) (0.46,0.613379) (0.47,0.616264) (0.48,0.619145) (0.49,0.622023) (0.5,0.624897) (0.51,0.627769) (0.52,0.630639) (0.53,0.633507) (0.54,0.636374) (0.55,0.639244) (0.56,0.64218) (0.57,0.645115) (0.58,0.64805) (0.59,0.650984) (0.6,0.653918) (0.61,0.656852) (0.62,0.659787) (0.63,0.662723) (0.64,0.66566) (0.65,0.668599) (0.66,0.67154) (0.67,0.674483) (0.68,0.677429) (0.69,0.680377) (0.7,0.683329) (0.71,0.686284) (0.72,0.689242) (0.73,0.692204) (0.74,0.695171) (0.75,0.698142) (0.76,0.701117) (0.77,0.704097) (0.78,0.707082) (0.79,0.710072) (0.8,0.713068)
};

\end{axis}
\end{tikzpicture}
\caption{The throughput of a traditional communication system is compared to that of an adaptive system with a switching policy that triggers an antenna reconfiguration whenever the channel is in state $1$.
As channel memory increases, the performance of the reconfigurable system surpasses the maximum service rate of the traditional implementation.}
\label{figure:ThroughputChannelMemory}
\end{center}
\end{figure}

Figure~\ref{figure:ThroughputChannelMemory} displays maximum throughput in bits per channel use as a function of the memory coefficient defined in \eqref{equation:ChannelMemoryParameter}.
When channel memory is small, the communication system with a fixed antenna structure performs better.
In particular, if the mixing time of the Gilbert-Elliott channel is shorter than a codeword transmission cycle, then reconfiguration offers little rewards.
It is therefore more profitable to send codewords constantly.
On the other hand, as the memory coefficient approaches one, the channel can get stuck in a bad fade for a prolonged period of time.
This phenomenon almost certainly guarantees decoding failure at the next attempt.
This prompts the RF front-end to trigger a reconfiguration and seek a more auspicious channel realization.
The crossover point in Fig.~\ref{figure:ThroughputChannelMemory} is approximately $0.28$.
Interestingly, at this crossover point, the expected sojourn time in state $1$ is approximately $113$~bits, which is very close to the actual block length.

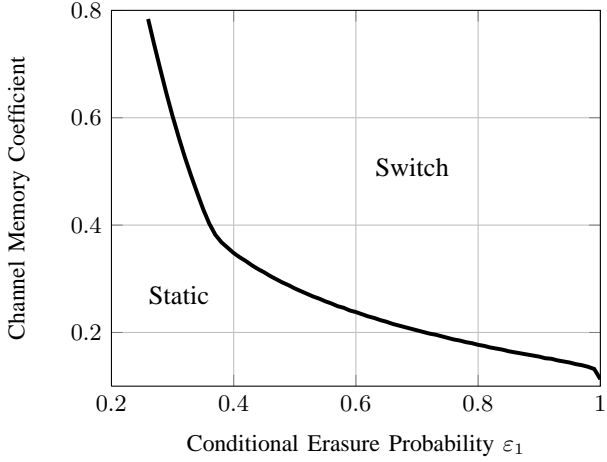
\begin{figure}[tb!]
\begin{center}
\setlength\tikzheight{5cm} 
\setlength\tikzwidth{6.5cm} 
%
%
\begin{tikzpicture}

\begin{axis}[%
font=\small,
scale only axis,
width=\tikzwidth,
height=\tikzheight,
xmin=0.2, xmax=1,
ymin=0.1, ymax=0.8,
xlabel={Conditional Erasure Probability $\varepsilon_1$},
ylabel={Channel Memory Coefficient},
xmajorgrids,
ymajorgrids]
\addplot [
color=black,
solid,
line width=1.5pt
]
coordinates{ (0.26,0.784) (0.27,0.736) (0.28,0.69) (0.29,0.645) (0.3,0.603) (0.31,0.564) (0.32,0.527) (0.33,0.493) (0.34,0.461) (0.35,0.43) (0.36,0.403) (0.37,0.382) (0.38,0.368) (0.39,0.358) (0.4,0.348) (0.41,0.34) (0.42,0.333) (0.43,0.325) (0.44,0.318) (0.45,0.312) (0.46,0.305) (0.47,0.299) (0.48,0.293) (0.49,0.288) (0.5,0.282) (0.51,0.277) (0.52,0.272) (0.53,0.267) (0.54,0.263) (0.55,0.258) (0.56,0.254) (0.57,0.249) (0.58,0.246) (0.59,0.241) (0.6,0.238) (0.61,0.234) (0.62,0.23) (0.63,0.227) (0.64,0.223) (0.65,0.22) (0.66,0.216) (0.67,0.213) (0.68,0.21) (0.69,0.207) (0.7,0.204) (0.71,0.201) (0.72,0.198) (0.73,0.196) (0.74,0.193) (0.75,0.19) (0.76,0.187) (0.77,0.185) (0.78,0.182) (0.79,0.18) (0.8,0.177) (0.81,0.175) (0.82,0.172) (0.83,0.17) (0.84,0.168) (0.85,0.165) (0.86,0.163) (0.87,0.161) (0.88,0.159) (0.89,0.157) (0.9,0.155) (0.91,0.152) (0.92,0.151) (0.93,0.148) (0.94,0.146) (0.95,0.144) (0.96,0.141) (0.97,0.139) (0.98,0.136) (0.99,0.132) (1,0.113)
};

\end{axis}

\pgftext[base,x=0.9cm,y=1.1cm] {Static}
\pgftext[base,x=4cm,y=2.8cm] {Switch}
\end{tikzpicture}
\caption{
System parameters determine which implementation delivers better throughput, the static system with a fixed antenna structure (Static) or the switching scheme with reconfigurable antennas (Switch).
Correlation and fade differentiation are advantageous to the RF-agile switching scheme.
In this figure, the probability of erasure is set at $0.20$ and $\varepsilon_2 = (1 - \varepsilon_1)/4$.
}
\label{figure:RegionBoundary}
\end{center}
\end{figure}

Similar curves can be generated for other parameter sets.
Each of these curves identifies a crossover point in terms of channel memory and erasure probability where the throughput of the adaptive system overtakes the expected service rate of the traditional implementation.
Plotting these points delineates the boundary of two regions, one where a static system with a fixed antenna structure performs better and a second region where the adaptive system delivers enhanced performance.
This is illustrated in Fig.~\ref{figure:RegionBoundary}.
We immediately see from this figure that channel correlation over time favors reconfigurable antennas.
In addition, experiencing vastly different channel qualities over the various fade levels benefits adaptive implementations.
Altogether, the capabilities of reconfigurable antenna systems seem better suited to harsh wireless environments.

\begin{figure}[tb!]
\begin{center}
\setlength\tikzheight{5cm} 
\setlength\tikzwidth{6.5cm} 
%
%
\begin{tikzpicture}

\begin{axis}[%
font=\small,
scale only axis,
width=\tikzwidth,
height=\tikzheight,
xmin=0, xmax=0.8,
ymin=6, ymax=22,
xlabel={Channel Memory Coefficient},
ylabel={Expected Waiting Time},
xmajorgrids,
ymajorgrids,
legend entries={Fixed Antenna, Switching Scheme},
legend style={nodes=right}]
\addplot [
color=black,
solid,
line width=1.5pt
]
coordinates{ (0,6.10476) (0.001,8.51002) (0.002,8.73542) (0.003,8.8859) (0.004,9.00235) (0.005,9.09881) (0.006,9.18191) (0.007,9.2554) (0.008,9.32157) (0.009,9.38197) (0.01,9.43768) (0.01,9.43768) (0.02,9.84857) (0.03,10.1304) (0.04,10.3523) (0.05,10.6342) (0.06,10.789) (0.07,10.9261) (0.08,11.1605) (0.09,11.2642) (0.1,11.3593) (0.11,11.5676) (0.12,11.639) (0.13,11.7055) (0.14,11.8931) (0.15,11.9401) (0.16,11.9842) (0.17,12.1523) (0.18,12.1794) (0.19,12.3425) (0.2,12.3533) (0.21,12.3632) (0.22,12.5065) (0.23,12.5017) (0.24,12.6406) (0.25,12.6217) (0.26,12.6033) (0.27,12.7269) (0.28,12.6957) (0.29,12.6657) (0.3,12.7817) (0.31,12.7402) (0.32,12.7003) (0.33,12.6622) (0.34,12.7722) (0.35,12.7242) (0.36,12.6783) (0.37,12.6346) (0.38,12.5929) (0.39,12.5533) (0.4,12.6612) (0.41,12.614) (0.42,12.5693) (0.43,12.5269) (0.44,12.4869) (0.45,12.4493) (0.46,12.4141) (0.47,12.3812) (0.48,12.3507) (0.49,12.3226) (0.5,12.4395) (0.51,12.4071) (0.52,12.3774) (0.53,12.3506) (0.54,12.3267) (0.55,12.3056) (0.56,12.2875) (0.57,12.2725) (0.58,12.2607) (0.59,12.2521) (0.6,12.2469) (0.61,12.2452) (0.62,12.2472) (0.63,12.2531) (0.64,12.263) (0.65,12.2773) (0.66,12.2963) (0.67,12.3201) (0.68,12.3492) (0.69,12.384) (0.7,12.425) (0.71,12.6031) (0.72,12.6526) (0.73,12.7102) (0.74,12.7768) (0.75,12.8533) (0.76,12.9409) (0.77,13.0408) (0.78,13.1547) (0.79,13.2844) (0.8,13.4321)
};

\addplot [
color=blue,
dashed,
line width=1.5pt
]
coordinates{ (0,12.2846) (0.001,19.7145) (0.002,20.2215) (0.003,20.5168) (0.004,20.7191) (0.005,20.868) (0.006,20.982) (0.007,21.0713) (0.008,21.1422) (0.009,21.1989) (0.01,21.2441) (0.01,21.2441) (0.02,21.6526) (0.03,21.4126) (0.04,21.3647) (0.05,20.928) (0.06,20.753) (0.07,20.5729) (0.08,20.0252) (0.09,19.805) (0.1,19.2548) (0.11,19.0183) (0.12,18.4787) (0.13,17.9752) (0.14,17.7208) (0.15,17.2339) (0.16,17.0015) (0.17,16.5333) (0.18,16.0951) (0.19,15.6838) (0.2,15.4724) (0.21,15.0772) (0.22,14.7052) (0.23,14.3539) (0.24,14.0216) (0.25,13.8562) (0.26,13.5357) (0.27,13.2316) (0.28,12.9427) (0.29,12.6677) (0.3,12.4054) (0.31,12.1549) (0.32,11.9153) (0.33,11.6858) (0.34,11.5795) (0.35,11.3561) (0.36,11.1419) (0.37,10.9362) (0.38,10.7385) (0.39,10.5482) (0.4,10.3649) (0.41,10.1882) (0.42,10.0176) (0.43,9.85275) (0.44,9.6934) (0.45,9.53919) (0.46,9.38986) (0.47,9.24514) (0.48,9.10479) (0.49,8.96858) (0.5,8.83631) (0.51,8.70779) (0.52,8.58283) (0.53,8.46127) (0.54,8.34294) (0.55,8.29399) (0.56,8.17748) (0.57,8.06402) (0.58,7.95349) (0.59,7.84574) (0.6,7.74067) (0.61,7.63815) (0.62,7.53809) (0.63,7.44037) (0.64,7.34491) (0.65,7.25161) (0.66,7.16038) (0.67,7.07116) (0.68,6.98385) (0.69,6.8984) (0.7,6.81472) (0.71,6.73277) (0.72,6.65246) (0.73,6.57376) (0.74,6.49659) (0.75,6.42091) (0.76,6.34666) (0.77,6.2738) (0.78,6.20229) (0.79,6.13207) (0.8,6.06311)
};

\end{axis}
\end{tikzpicture}
\caption{Mean waiting times for traditional and switching systems are plotted as functions of channel memory;
a smaller waiting time is desirable.
When the channel is weakly correlated over time, the system with a fixed antenna configuration performs better.
On the other hand, in slow fading scenarios, the adaptive implementation with a reconfigurable antenna structure becomes advantageous.}
\label{figure:MeanWaitingTime}
\end{center}
\end{figure}
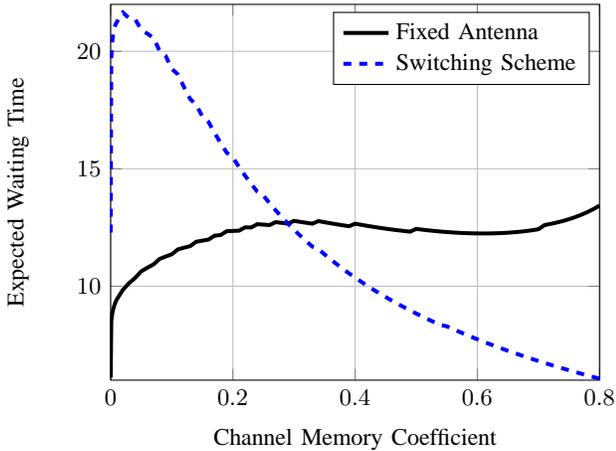

We supplement the preceding results by investigating the performance of the two competing systems in relation to mean waiting time.
This latter criterion is appropriate for lightly loaded connections and delay-sensitive applications such as mobile telephony and video conferencing.
Again, we maintain the average rate of bit erasure at 20 percent, and we set the conditional probabilities of erasure to $\varepsilon_1 = 0.5$ and $\varepsilon_2 = 0.125$.
As before, we vary the channel memory coefficient to first produce a memoryless process followed by increasingly correlated erasure sequences.
Figure~\ref{figure:MeanWaitingTime} plots the mean waiting time at the source as a function of memory.
The crossover point where the switching system with a reconfigurable antenna structure overtakes the static implementation is approximately the same as in the case of throughput.
In fact, preliminary results indicate that similar behavior can be observed for various parameter sets and different optimization criteria including mean waiting time, asymptotic decay rate in queue occupancy and threshold violation probability.
This robustness may be attributable to the simplicity of the Gilbert-Elliott model and may not hold for more complex channel models.
This observation warrants further research.
In practice, this suggests that good performance can be achieved with RF-agile antenna structures by identifying regions where reconfiguration should take place.
The system can then estimate the current state of the channel and decide, according to its local map, whether or not a reconfiguration event should be triggered.

\subsection{Rayleigh Fading Approximation by Finite-State Channel}

In this section, we turn our attention to an 8-state channel derived from the Rayleigh fading model.
In a landmark article~\cite{Wang1995tvt}, the authors describe a structured methodology to construct finite-state Markov models from Rayleigh channels.
We refer the reader to \cite{Wang1995tvt,Zhang1999tcom} for the details.
The parameters we select for the model are the following: average SNR of $-5$ dB, at a transmission rate of $10^5$ bits per second.
For the 8-state channel, the erasure probabilities $\{ \varepsilon_i \}$ are given by $0.4244$, $0.3591$, $0.3134$, $0.2732$, $0.2348$, $0.1954$, $0.1512$, $0.0879$ for $i = 1, 2, \ldots, 8$, respectively.
This gives an average erasure rate of $0.2549$.
The remaining parameters in our system model are the same as before: $N=114$, $\gamma=0.2$, $\rho^{-1}=195$.

\begin{figure}
\begin{center}
\setlength\tikzheight{5cm} 
\setlength\tikzwidth{6.5cm} 
%
%
\begin{tikzpicture}

\definecolor{mygreen}{rgb}{0,0.4,0}

\begin{axis}[%
font=\small,
scale only axis,
width=\tikzwidth,
height=\tikzheight,
xmin=40, xmax=120,
ymin=0.5, ymax=0.64,
ylabel={Throughput (bits per channel use)},
xlabel={Doppler Frequency (Hz)},
xmajorgrids,
ymajorgrids,
zmajorgrids,
]

\addplot [
color=black,
solid,
line width=1.5pt,
]
coordinates{
 (40,0.526109)(45,0.526327)(50,0.526544)(55,0.526759)(60,0.526972)(65,0.527184)(70,0.527394)(75,0.527603)(80,0.52781)(85,0.528016)(90,0.52822)(95,0.528422)(100,0.528623)(105,0.528823)(110,0.529021)(115,0.529218)(120,0.529413) 
};

\addplot [
color=red,
loosely dashed,
line width=1.5pt,
]
coordinates{
 (40,0.630626)(45,0.626585)(50,0.62265)(55,0.618816)(60,0.615079)(65,0.611433)(70,0.607876)(75,0.604402)(80,0.601009)(85,0.597692)(90,0.594527)(95,0.59144)(100,0.588419)(105,0.585461)(110,0.582562)(115,0.579722)(120,0.576937) 
};

\addplot [
color=blue,
dashed,
line width=1.5pt,
]
coordinates{
 (40,0.638624)(45,0.632496)(50,0.626582)(55,0.620869)(60,0.615346)(65,0.610001)(70,0.604823)(75,0.599804)(80,0.595006)(85,0.59039)(90,0.585905)(95,0.581544)(100,0.5773)(105,0.573169)(110,0.569144)(115,0.56522)(120,0.561394) 
};

\addplot [
color=mygreen,
densely dashed,
line width=1.5pt,
]
coordinates{
 (40,0.626505)(45,0.616795)(50,0.607566)(55,0.59878)(60,0.590402)(65,0.582401)(70,0.57475)(75,0.567423)(80,0.560398)(85,0.553745)(90,0.547387)(95,0.541269)(100,0.535376)(105,0.529695)(110,0.524212)(115,0.518916)(120,0.513796) 
};

\end{axis}
\end{tikzpicture}
\caption{Throughput of the standard and reconfigurable systems for an 8-state Markov model derived from a Rayleigh fading channel.
The solid line corresponds to the fixed antenna system; and the dashed lines, to reconfigurable antenna systems with different switching policies.
The switching sets for the loosely dashed red line, dashed blue line, and densely dashed green line are $\{ 1, 2, 3 \}$, $\{ 1, 2, 3, 4 \}$ and $\{ 1, 2, 3, 4, 5 \}$, respectively.
A small Doppler frequency characterizes a slow fading channel.
}
\label{figure:TPut_Rayleigh_8State}
\end{center}
\end{figure}
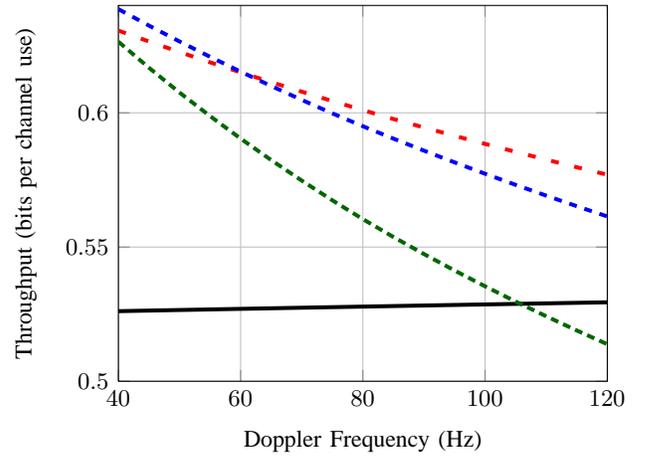

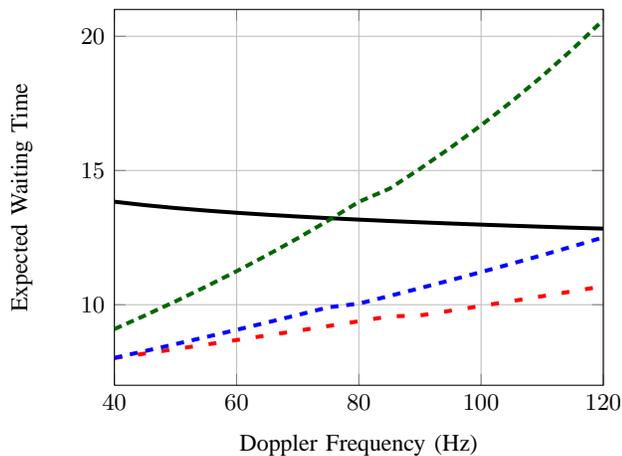
\begin{figure}[ht!]
\begin{center}
\setlength\tikzheight{5cm} 
\setlength\tikzwidth{6.5cm} 
%
%
\begin{tikzpicture}

\definecolor{mygreen}{rgb}{0,0.4,0}

\begin{axis}[%
font=\small,
scale only axis,
width=\tikzwidth,
height=\tikzheight,
xmin=40, xmax=120,
ymin=7, ymax=21,
xmajorgrids,
ymajorgrids,
zmajorgrids,
xlabel={Doppler Frequency (Hz)},
ylabel={Expected Waiting Time},
]

\addplot [
color=black,
solid,
line width=1.5pt,
]
coordinates{
 (40,13.8373)(45,13.7094)(50,13.6012)(55,13.5077)(60,13.4254)(65,13.3522)(70,13.286)(75,13.2257)(80,13.1703)(85,13.119)(90,13.0711)(95,13.0262)(100,12.9839)(105,12.9439)(110,12.9058)(115,12.8696)(120,12.8349) 
};

\addplot [
color=red,
loosely dashed,
line width=1.5pt,
]
coordinates{
 (40,8.01622)(45,8.18075)(50,8.34708)(55,8.5152)(60,8.68513)(65,8.8569)(70,9.03058)(75,9.20623)(80,9.38392)(85,9.56373)(90,9.59986)(95,9.77584)(100,9.95389)(105,10.1341)(110,10.3165)(115,10.5013)(120,10.6884) 
};

\addplot [
color=blue,
dashed,
line width=1.5pt,
]
coordinates{
 (40,8.01884)(45,8.27387)(50,8.53312)(55,8.79687)(60,9.06541)(65,9.33905)(70,9.61812)(75,9.90294)(80,10.0411)(85,10.3257)(90,10.6164)(95,10.9135)(100,11.2174)(105,11.5285)(110,11.8471)(115,12.1738)(120,12.5089) 
};

\addplot [
color=mygreen,
densely dashed,
line width=1.5pt,
]
coordinates{
 (40,9.09258)(45,9.60003)(50,10.1273)(55,10.6764)(60,11.2495)(65,11.849)(70,12.4775)(75,13.1381)(80,13.8338)(85,14.321)(90,15.0641)(95,15.8502)(100,16.684)(105,17.5704)(110,18.5155)(115,19.5259)(120,20.6093) 
};

\end{axis}
\end{tikzpicture}
\caption{Mean waiting time of the standard and reconfigurable systems for an 8-state Markov model derived from a Rayleigh fading channel.
The solid line is associated with the fixed antenna system.
The dashed lines represent the performance of the reconfigurable systems described in the previous figure.}
\label{figure:ExpWaitTime_Rayleigh_8State}
\end{center}
\end{figure}

We study the performance of our two competing systems by varying the Doppler frequency in the Rayleigh fading channel model.
Figures~\ref{figure:TPut_Rayleigh_8State} and \ref{figure:ExpWaitTime_Rayleigh_8State} show the throughput and average waiting time, respectively, as a function of the Doppler frequency.
The solid line is associated with the fixed antenna system.
The dashed lines correspond to the reconfigurable antenna system with different switching policies.
The loosely dashed red line corresponds to the policy that triggers an antenna reconfiguration when the channel is in states $\{ 1, 2, 3 \}$, while the dashed blue and densely dashed green lines correspond to $\{1,2,3,4\}$, $\{1,2,3,4,5\}$, respectively.
We note that a slow fading channel corresponds to a small Doppler frequency.
The key insights obtained from Gilbert-Elliott channel appear to hold here as well, that is, in a slow fading channel, having a reconfigurable antenna counteracts the adverse effects of a deep fade. 

\subsection{POMDP Formulation}

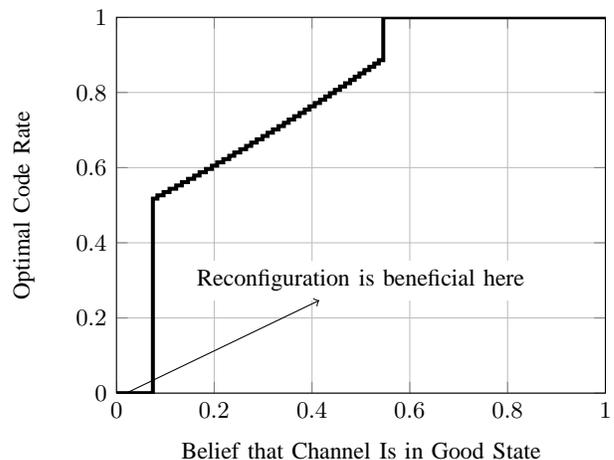
\begin{figure}[tb!]
\begin{center}
\setlength\tikzheight{5cm} 
\setlength\tikzwidth{6.5cm} 
%
%
\begin{tikzpicture}

\begin{axis}[%
font=\small,
scale only axis,
width=\tikzwidth,
height=\tikzheight,
xmin=0, xmax=1,
ymin=0, ymax=1,
xlabel={Belief that Channel Is in Good State},
ylabel={Optimal Code Rate},
xmajorgrids,
ymajorgrids,
zmajorgrids]
\addplot [
color=black,
solid,
line width=1.5pt
]
coordinates{
(0,0)(0.0745515,0)(0.0745515,0.517544)(0.0838071,0.517544)(0.0838071,0.526316)(0.0964251,0.526316)(0.0964251,0.535088)(0.108616,0.535088)(0.108616,0.54386)(0.122213,0.54386)(0.122213,0.552632)(0.133766,0.552632)(0.133766,0.561404)(0.146396,0.561404)(0.146396,0.570175)(0.158982,0.570175)(0.158982,0.578947)(0.170063,0.578947)(0.170063,0.587719)(0.182978,0.587719)(0.182978,0.596491)(0.195209,0.596491)(0.195209,0.605263)(0.20601,0.605263)(0.20601,0.614035)(0.218547,0.614035)(0.218547,0.622807)(0.230712,0.622807)(0.230712,0.631579)(0.240993,0.631579)(0.240993,0.640351)(0.25318,0.640351)(0.25318,0.649123)(0.265244,0.649123)(0.265244,0.657895)(0.275645,0.657895)(0.275645,0.666667)(0.286951,0.666667)(0.286951,0.675439)(0.2987,0.675439)(0.2987,0.684211)(0.309147,0.684211)(0.309147,0.692982)(0.320008,0.692982)(0.320008,0.701754)(0.331466,0.701754)(0.331466,0.710526)(0.342473,0.710526)(0.342473,0.719298)(0.352257,0.719298)(0.352257,0.72807)(0.363448,0.72807)(0.363448,0.736842)(0.374546,0.736842)(0.374546,0.745614)(0.385114,0.745614)(0.385114,0.754386)(0.394714,0.754386)(0.394714,0.763158)(0.405575,0.763158)(0.405575,0.77193)(0.416352,0.77193)(0.416352,0.780702)(0.427048,0.780702)(0.427048,0.789474)(0.436597,0.789474)(0.436597,0.798246)(0.446696,0.798246)(0.446696,0.807018)(0.457192,0.807018)(0.457192,0.815789)(0.467617,0.815789)(0.467617,0.824561)(0.477972,0.824561)(0.477972,0.833333)(0.487312,0.833333)(0.487312,0.842105)(0.49701,0.842105)(0.49701,0.850877)(0.507178,0.850877)(0.507178,0.859649)(0.517268,0.859649)(0.517268,0.868421)(0.527263,0.868421)(0.527263,0.877193)(0.536672,0.877193)(0.536672,0.885965)(0.545713,0.885965)(0.545713,1)(1,1)
};

\node[fill=white] (note1) at (axis description cs:0.5,0.3) {Reconfiguration is beneficial here};
\draw[->] (axis description cs:0.02,0.0) -- (note1);
\end{axis}
\end{tikzpicture}
\caption{
This figure shows the choice of optimal code rate versus the belief that the channel is in good state.
Consistent with intuition, when the transmitter is more confident that the channel is in good state, it can use a high rate code.
Conversely, when the transmitter is certain that the channel is in bad state, reconfiguring the antenna is beneficial.
We note that the optimal code rate is a threshold based policy, partitioning the belief space.
}
\label{figure:POMDP_2state}
\end{center}
\end{figure}

\begin{figure}[tb!]
\begin{center}
\setlength\tikzheight{5cm} 
\setlength\tikzwidth{6.5cm} 
\input{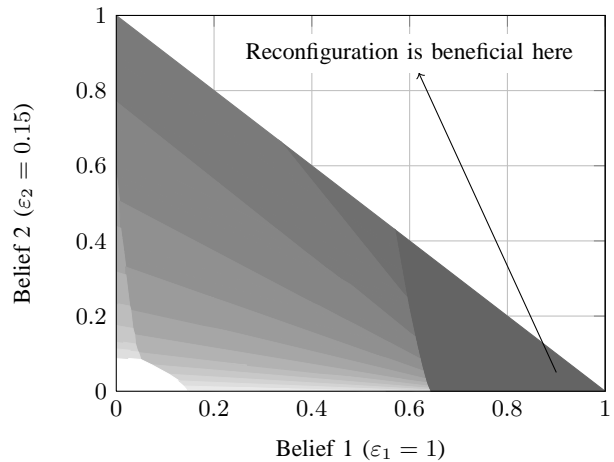}
\caption{This figure shows the choice of optimal code rate according to the belief.
The darkest region corresponds to the belief space where a reconfiguration event is beneficial.
The region in white corresponds to the belief space where using a code rate of $1$ is beneficial.
The regions in gray correspond to different code rates; lighter regions map to higher rates.}
\label{figure:POMDP_3state}
\end{center}
\end{figure}

Figure~\ref{figure:POMDP_2state} shows the numerical analysis of a two-state channel from the POMDP perspective presented in Section \ref{subsection:POMDPPerspective}.
The parameters for the analysis are: $\varepsilon_1=1$, $\varepsilon_2=0$, a channel memory coefficient of $0.3$, an average erasure rate of 20\%, and discount factor $\beta=0.9$.
For this two-state channel, the belief vector lies in a two-dimensional simplex.
Yet, since the two elements in the belief vector sum to one, only the second coefficient is kept, namely the belief that the channel is in a good state.
More precisely, Fig.~\ref{figure:POMDP_2state} displays the decision regions (optimal action to choose, i.e., $\delta^{*}$) as a function of the belief that the channel is in a good state.
We observe that the optimal policy is a threshold based policy partitioning the belief space.
The optimal code rate is monotonically increasing as a function of belief state.
Figure~\ref{figure:POMDP_3state} shows a similar analysis for a three-state channel with a transition probability matrix
\begin{align*} 
\mathbf{B} = \begin{bmatrix}
0.998 & 0.002 & 0 \\
0.001 & 0.998 & 0.001 \\
0 & 0.002 & 0.998
\end{bmatrix} ,
\end{align*}
and erasure values $\varepsilon_1=1$, $\varepsilon_2=0.15$, $\varepsilon_3=0$.
This gives an average erasure rate of 20\% and a channel memory coefficient of $0.8$.
The discount factor used for the computation of the infinite horizon value function is $\beta=0.9$.

\begin{figure}[tb!]
\begin{center}
\setlength\tikzheight{5cm} 
\setlength\tikzwidth{6.5cm} 
%
%

\begin{tikzpicture}

\begin{axis}[%
font=\small,
scale only axis,
width=\tikzwidth,
height=\tikzheight,
xmin=0.1, xmax=0.9,
ymin=2, ymax=9,
xlabel={Channel Memory Coefficient},
ylabel={Mean Value Function},
xmajorgrids,
ymajorgrids,
zmajorgrids,
legend entries={Fixed Antenna,Reconfigurable Antenna},
legend style={at={(0.97,0.03)},anchor=south east,nodes=right}]]
\addplot [
color=black,
solid,
line width=1.5pt
]
coordinates{
 (0.1,5.08181)(0.15,5.33837)(0.2,5.52839)(0.25,5.68756)(0.3,5.8067)(0.35,5.89968)(0.4,5.96839)(0.45,6.02465)(0.5,6.04746)(0.55,6.04106)(0.6,6.00921)(0.65,5.93015)(0.7,5.80824)(0.75,5.62613)(0.8,5.3026)(0.85,4.80827)(0.9,4.00926) 
};

\addplot [
color=blue,
dashed,
line width=1.5pt
]
coordinates{
 (0.1,5.08181)(0.15,5.33837)(0.2,5.52839)(0.25,5.68756)(0.3,5.8067)(0.35,5.90014)(0.4,6.28625)(0.45,6.50643)(0.5,6.73286)(0.55,7.00034)(0.6,7.26049)(0.65,7.51528)(0.7,7.7707)(0.75,8.02166)(0.8,8.26724)(0.85,8.51288)(0.9,8.76976) 
};

\end{axis}
\end{tikzpicture}
\caption{Comparing the mean value functions for the traditional and reconfigurable antenna systems.
The greater flexibility of the adaptive system yields a dominating performance curve.
Adaptation pays off in harsh environments where performance is strictly better.}
\label{figure:MeanValueFunction_2state}
\end{center}
\end{figure}
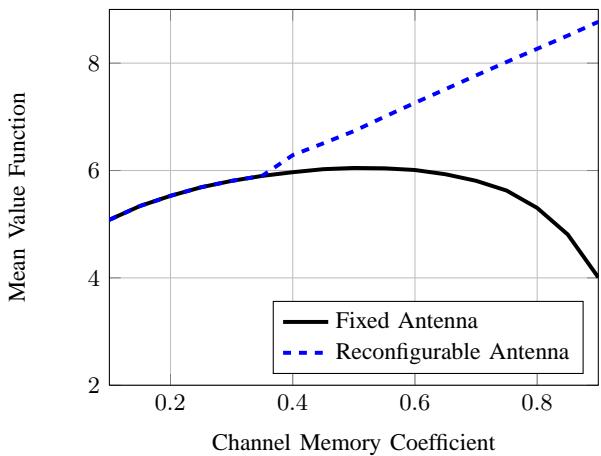

Figure~\ref{figure:MeanValueFunction_2state} plots the expected value function averaged over the belief space for a two-state channel with $\varepsilon_1=1$, $\varepsilon_2=0$, an average erasure rate of 20\%, and discount factor $\beta=0.9$.
In the high memory regime, whenever the channel is in bad state, successful receptions in the fixed antenna system are heavily delayed.
Since, the delayed rewards are discounted by a factor of $\beta=0.9$, the average value function is small in this regime for this system.
On the contrary, in the reconfigurable antenna system, prolonged bad channel states can be circumvented by issuing a reconfiguration event and thereafter transmitting during the prolonged good channel states, thus exploiting the benefits of high channel memory. 
This explains the large gap between the mean value functions of the two systems in this regime.    

\vspace{-0.3cm}
\section{Concluding Remarks}
\label{section:ConcludingRemarks}

This preliminary study offers supporting evidence to the claim that reconfigurable antenna structures can improve the performance of communication systems significantly.
For the reconfiguration process to be beneficial, the potential rewards of a reconfiguration event must offset the costs of a loss of a codeword transmission opportunity.
Two conditions appear to influence this balance.
The coherence time of the physical channel must be on the order of the codeword cycle or longer.
Furthermore, the quality of the channel must vary significantly over the different fade levels.
Slow fading channels appear to be great prospects for reconfigurable antenna systems with adaptive control policies.

Future studies should address practical issues such as complexity and power efficiency.
Once side information becomes available at the transmitter, power control and scheduling can be employed in conjunction with reconfigurable antenna structures.
Extending the queueing formulation to account for these techniques is an interesting goal.
Also, the postulate that virtual channels are independent from one another should be explored through empirical measurements.
A strong positive correlation among virtual channels could reduce the expected returns of a reconfiguration event.
These are promising avenues of future research that may broaden the application potential of reconfigurable antennas and help improve the performance of wireless communication systems.

\end{document}